\documentclass[aps,prb, superscriptaddress,
 amsmath,amssymb, 
 reprint,%
]{revtex4-2}

\usepackage{graphicx}
\usepackage{subfigure}
\usepackage{dcolumn}
\usepackage{multirow}
\usepackage{bm}
\usepackage{array}
\usepackage[colorlinks=true,linkcolor=black,anchorcolor=black,citecolor=black,urlcolor=black]{hyperref}
\usepackage{comment}
\usepackage[table,xcdraw]{xcolor}
\usepackage{float}
\usepackage{longtable}
\usepackage{mathptmx}
\usepackage{amsmath}
\usepackage{etoolbox, xcolor}
\usepackage{soul}
\usepackage{amssymb}
\usepackage{pifont}

\makeatletter
\def\ignorecitefornumbering#1{%
     \begingroup
         \@fileswfalse
         #1
    \endgroup
}
\makeatother

\begin{document}



\title{Zero-Dipole Schottky Contact: Homologous Metal Contact to 2D Semiconductor}

\author{Che Chen Tho}
\affiliation{ 
Science, Mathematics and Technology Cluster, Singapore University of Technology and Design, Singapore 487372}%

\author{Yee Sin Ang}
\email{yeesin\_ang@sutd.edu.sg}
\affiliation{ 
Science, Mathematics and Technology Cluster, Singapore University of Technology and Design, Singapore 487372}%

\begin{abstract}

Band alignment of metal contacts to 2D semiconductors often deviate from the ideal Shottky-Mott (SM) rule due to the non-ideal factors such as the formation of interface dipole and metal-induced gap states (MIGS). Although MIGS can be strongly suppressed using van der Waals (vdW) contact engineering, the interface dipole is hard to eliminate due to the electronegativity difference of the two contacting materials. Here we show that interface dipole can be practically eliminated in 2D semiconducting MoSi$_2$N$_4$ when contacted by its homologous metallic counterpart MoSi$_2$N$_4$(MoN)$_n$ ($n = 1-4$). The SiN outer sublayers, simultaneously present in both MoSi$_2$N$_4$ and MoSi$_2$N$_4$(MoN)$_n$, creates nearly equal charge `push-back' effect at the contact interface. This nearly symmetrical charge redistribution leads to zero net electron transfer across the interface, resulting in a \emph{zero-dipole} contact. 
Intriguingly, we show that even in the extreme close-contact case where MoSi$_2$N$_4$(MoN) is arbitrarily pushed towards MoSi$_2$N$_4$ with extremely small interlayer distance, the interface dipole remains practically zero. 
Such \emph{zero-dipole} Schottky contact represents a peculiar case where the SM rule, usually expected to occur only in the non-interacting regime, manifests in MoSi$_2$N$_4$/MoSi$_2$N$_4$(MoN)$_n$ vdWH even though the constituent monolayers interact strongly. A model for pressure sensing is then proposed based on changing the interlayer distance in MoSi$_2$N$_4$/MoSi$_2$N$_4$(MoN) vdWH. 

\end{abstract}

\maketitle

\section{\label{sec: introduction}Introduction}

\begin{figure*}
\centering
\includegraphics[width=0.85\textwidth]{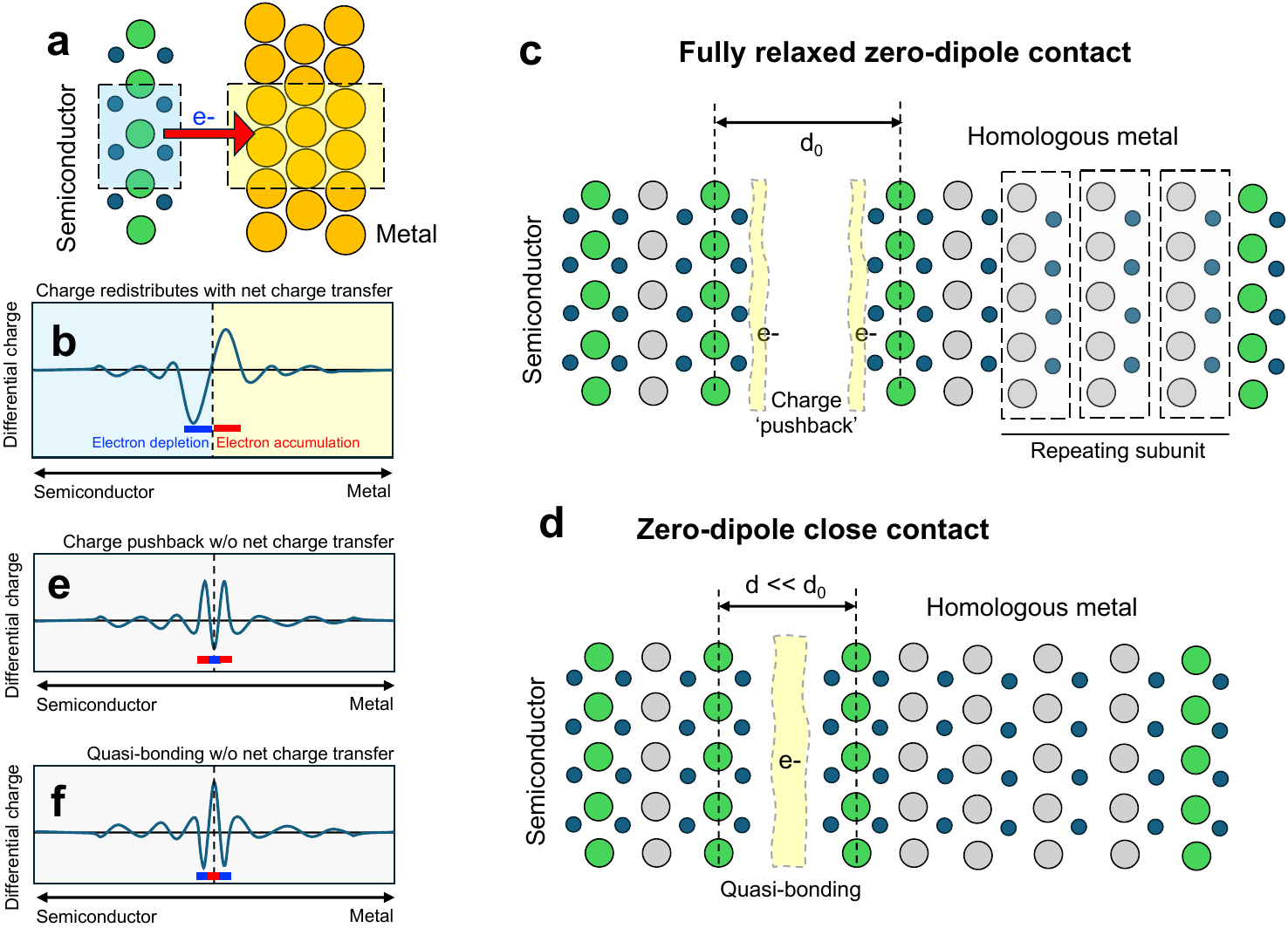}
\caption{\label{Fig1}\textbf{Concept of \emph{zero-dipole} metal-semiconductor (MS) contact.} (a) Charge redistribution leads to net charge transfer across an MS contact. (b) The differential charge diagram in which electrons are depleted at the semiconductor but accumulate at the metal. (c) MS contact composed of a semiconductor and its homologous metal. The similar atomic layers at the contacting interface leads to equal charge push-back effect without net charge transfer across the contact. (d) In the close-contact regime, \emph{quasi-bonding} occurs but the net charge transfer is still practically zero. (e) and (f) Differential charge density profile of (c) and (d), respectively. }
\end{figure*}

Two-dimensional (2D) semiconductors are promising materials for transistors that are ubiquitous in modern electronics \cite{transistor_2D_promises, transistor_2D_IC, transistor_2D_vertical_promises, transistor_2D_new_mechanism}. 
Comparing to bulk (3D) semiconductors such as silicon, the atomically smooth interface free of dangling bonds and ultrathin body of 2D semiconductors \cite{transistor_2D_new_mechanism, 2D_material_transistor, chhowalla2016_2D_transistor, 2D_material_gate_control, layered_2D_material_transistor} are beneficial for designing high-performance field-effect transistor (FET) \cite{chhowalla2016_2D_transistor,transistor_2D_promises} with significantly improved electrostatic control in sub-10-nm gate-length regime \cite{jinglu2021_schottky}. 2D semiconductors are therefore a promising material platform to complement existing silicon-based technology or to realise innovations that are beyond it \cite{jayachandran2024_integration_3D_2D, kim2024_beyond_Moore_law, huang2022_2D_semiconductor_eletronic_application, su2023_Janus}. 
Transforming 2D semiconductors into industrial-grade device technology is, however, faced with many challenges \cite{li20192d} such as the growth of large-area crystal with high uniformity, and the fabrication of high-quality contact interfaces with dielectrics \cite{illarionov2020insulators, lau2023dielectrics} and metal electrodes \cite{allain2015electrical, schulman2018contact}.

A key challenge that has received tremendous research interests in the past decade \cite{ma2024van} when it comes to developing 2D semiconductor device technology, is the design of efficient 2D metal-semiconductor (MS) contacts. Charge transport across MS contacts is sensitively influenced by an energy barrier called the Schottky barrier height (SBH) which arises from the energy mismatch between the semiconductor band edges and the metal work function ($W_M$). SBH has the effect of impeding charge injection across the MS contact interface \cite{Schottky_Mott} as charge carriers with energies less than this barrier cannot be transmitted without relying on tunneling . At the ideal non-interacting limit, the SBH can be determined by the Schottky-Mott (SM) relation \cite{liu2018_Schottky_Mott_vdWH}:
\begin{equation}\label{n_SBH}
\Phi_\text{n} = W_{M}-\varepsilon_\text{CBM}
\end{equation}
\begin{equation}\label{p_SBH}
\Phi_\text{p} = \varepsilon_\text{VBM}-W_{M}
\end{equation}
where $\Phi_{n(p)}$ is the n(p)-type SBH, $\varepsilon_{CBM}$ and $\varepsilon_{VBM}$ is the energy of the conduction band minimum (CBM) and valence band maximum (VBM) of the semiconductor as measured from the vacuum energy level ($E_{vac}$), respectively. In the presence of interfacial interactions, strains and dielectric screening, the SBH can deviate substantially from the ideal SM rule \cite{Schottky_Mott, park2021_schottky_mott_dielectric_screening, briggs2019_2D_materials_roadmap, tersoff1984_quantum_dipole}. 
Although van der Waals heterostructures (vdWHs) \cite{geim2013_vdWH, novoselov2016_vdWH} contacts using both 2D \cite{liu2016_weak_FLP} and 3D metals \cite{liu2018_Schottky_Mott_vdWH, wang2019_vdw_contact} can recover the SM rule due to the weak interlayer coupling and the minimal metal-induced gap states (MIGS), the interfacial charge transfer across the contact [Fig. \ref{Fig1}(a) and (b)] can still form an interface dipole, resulting in the SBH to deviate from the SM rule \cite{li2022_metal_semiconductor_dipole, Transparent_Dipole}. 
Due to the presence of such interface dipole even in the weak coupling limit, the prediction and engineering of SBH in 2D MS contacts cannot be straightforwardly achieved.  

In this work, we show that the interface dipole can be suppressed when contacting a 2D semiconductor with its \emph{homologous metallic counterpart} as the electrode material. A homologous material, is based on the blueprint of a parent material, through the increments of a specific sub-unit \cite{kimizuka1988_homologous, li1998_homologous} [Fig. \ref{Fig1}(c)] in particular directions. Using MoSi$_2$N$_4$ \cite{Hong2020,Wang2021_MA2Z4,yin2023_review,Tho2023_review,novoselov2024_MA2Z4_review, wang2024_intercalating_charge_density_wave} and its homologous metals - MoSi$_2$N$_4$(MoN)$_n$ ($n = 1-4$) \cite{NSR, Tho2024_MA2Z4(M'Z)} whereby Mo-N sublayer is the repeating unit, we show that \emph{zero-dipole} contact can be formed at the MoSi$_2$N$_4$/MoSi$_2$N$_4$(MoN)$_n$ vdWHs. 
Such \emph{zero-dipole} MS contacts arise from the similar atomic species and configurations of the contacting surfaces, which leads to the suppression of charge transfer across the materials. 
Intriguingly, when the interlayer distance is reduced to the close-contact regime [Fig. \ref{Fig1}(d)], the interfacial charge transfer continuously evolves from the case of charge `push-back' effect [Fig. \ref{Fig1}(c) and (e)] \cite{bokdam2014_pushback} towards the \emph{quasi-bonding} case [Fig. \ref{Fig1}(f)], yet still maintaining a \emph{zero-dipole} interface despite the presence of strong interfacial interactions.
We identify MoSi$_2$N$_4$/MoSi$_2$N$_4$(MoN) with ultralow SBH of 0.06 eV that can be beneficial for electronic device applications operating at near room temperatures, which we further investigated its potential for use as a pressure sensor.
These findings reveal the potential role of using homologous metals to achieve SM rule at the close-contact regime for nanodevice applications.

\section{\label{sec:Methods}Computational Methods}

We use the Vienna Ab initio Simulation Package (VASP) \cite{Hafner2008} to perform first-principles calculations from density functional theory (DFT) \cite{Hohenberg_1964}. A vacuum thickness of 20 $\mathrm{\AA}$ is inserted at each end of the lattice structure to eliminate  spurious interactions with the periodic images, and the optB88-vdW functional is adopted to take into account the vdW forces. We choose the Perdew-Burke-Ernzerhof generalised gradient approximation (PBE-GGA) \cite{PBE} for the exchange-correlation functional in our structural optimization and electronic properties calculations, since band gap value calculated using PBE for MoSi$_2$N$_4$ is closer to the experimentally obtained value than using HSE06 \cite{Hong2020}. A Brillouin zone k-point sampling grid of 15 $\times$ 15 $\times$ 1 is used for all our structural relaxation. For self-consistent field calculations, a denser $\Gamma$-centered Brillouin zone k-point sampling grid of 19 $\times$ 19 $\times$ 1 is used. The break condition for each ionic relaxation loop is set to 10$^{-4}$ eV Å$^{-1}$ for the forces acting on the unit cell, whereas the break condition for each electronic relaxation loop is set to 10$^{-6}$ eV. The vdWHs are constructed using QuantumATK \cite{Quantumatk} without the use of supercells. We keep the lattice constant of MoSi$_2$N$_4$ fixed while strained only MoSi$_2$N$_4$(MoN)$_n$ (0.3-0.6\%) to preserve the semiconducting characteristics of MoSi$_2$N$_4$ \cite{Wang2021}. During structural relaxation, both the lattice constants and atomic positions are allowed to relax for the case of monolayers, whereas only the atomic positions are allowed to relax for vdWHs.

\begin{table*}[]
\caption{\label{Stability} Optimised lattice constants, thickness ($t$), interlayer distance ($d$), elastic constants ($C_{11}$, $C_{12}$, $C_{66}$), 2D and 3D Young's modulus ($Y^{2D}$, $Y^{3D}$), 3D bulk modulus ($K^{3D}$), 3D sheer modulus ($G^{3D}$), Poisson's ratio ($\nu$) and formation energy ($H^{sys}_{f}$) of the monolayers and van der Waals heterostructures (vdWHs).} 

\resizebox{\textwidth}{!}{%

\begin{tabular}{>{\centering\arraybackslash}m{4cm}>{\centering\arraybackslash}m{2cm}>{\centering\arraybackslash}m{1.5cm}>{\centering\arraybackslash}m{1.5cm}>{\centering\arraybackslash}m{1.5cm}>{\centering\arraybackslash}m{1.5cm}>{\centering\arraybackslash}m{1.5cm}>{\centering\arraybackslash}m{2.5cm}>{\centering\arraybackslash}m{2cm}>{\centering\arraybackslash}m{1.5cm}>{\centering\arraybackslash}m{1.5cm}>{\centering\arraybackslash}m{1.5cm}>{\centering\arraybackslash}m{1.5cm}}

\hline \hline  
\textbf{Materials} & \textbf{Lattice   Constant ($\AA$)} & \textbf{t ($\AA$)}& \textbf{d ($\AA$)} & \textbf{C$_{11}$ (N/m)} & \textbf{C$_{12}$ (N/m)} & \textbf{C$_{66}$ (N/m)} & \textbf{$Y^{2D}$ (N/m)} & \textbf{$Y^{3D}$ (GPa)} & \textbf{$K^{3D}$ (GPa)} & \textbf{$G^{3D}$ (GPa)} & \textbf{$\nu$} & \textbf{$H^{sys}_{f}$ (eV/atom)} \\ \hline

\emph{Monolayers}   \\ 

MoSi$_2$N$_4$ & 2.91 & 7.00 &--& 544 & 153 & 196 & 501 & 496 & 345 & 194 & 0.28 & -0.818\\ 

MoSi$_2$N$_4$(MoN) & 2.92 & 9.81 &--& 663 & 161 & 251 & 624 & 484 & 319 & 195 & 0.24 & -0.754\\ 

MoSi$_2$N$_4$(MoN)$_2$ & 2.90 & 12.73 &--& 783 & 231 & 276 & 715 & 452 & 323 & 174 & 0.30 & -0.208\\ 

MoSi$_2$N$_4$(MoN)$_3$ & 2.89 & 15.61 &--& 980 & 248 & 366 & 917 & 490 & 328 & 196 & 0.25 & -0.147\\ 

MoSi$_2$N$_4$(MoN)$_4$ & 2.89 & 18.45 &--& 1152 & 270  & 441 & 1089 & 506 & 328 & 206 & 0.23 & -0.791\\ \\

\emph{Heterostructures} \\

MoSi$_2$N$_4$ / MoSi$_2$N$_4$(MoN)  & 2.91 & 19.88 & 3.04 & 1233 & 303 & 465 & 1158 & 504 & 334  & 202 & 0.25 & -0.888\\ 

MoSi$_2$N$_4$ / MoSi$_2$N$_4$(MoN)$_2$ & 2.91 & 22.75 & 3.03 & 1311 & 368 & 471 & 1208 & 467 & 325 & 182 & 0.28 & -0.848\\ 

MoSi$_2$N$_4$ / MoSi$_2$N$_4$(MoN)$_3$ & 2.91 & 25.63 & 3.05 & 1461 & 397 & 532 & 1353 & 471 & 323 & 185 & 0.27  & -0.810\\ 

MoSi$_2$N$_4$ / MoSi$_2$N$_4$(MoN)$_4$ & 2.91 & 28.45 & 3.04 & 1645 & 417 & 614 & 1539 &  488 & 327 & 195 & 0.25  & -0.778\\ 

\hline \hline
\end{tabular}%
}
\end{table*}

\section{\label{sec:Results}Results and Discussions}

\subsection{\label{monolayer} MoSi$_2$N$_4$ and MoSi$_2$N$_4$(MoN)$_n$ monolayers}

In this section, we investigate the physical properties of MoSi$_2$N$_4$(MoN)$_n$ ($n = 1-4$) monolayers, which can be organised into three subsections: \emph{1. Structure and mechanical properties}; \emph{2. Stability}; and \emph{3. Electronic properties}.

\subsubsection{Structure and mechanical properties}
\begin{figure*}
\centering
\includegraphics[width=0.958\textwidth]{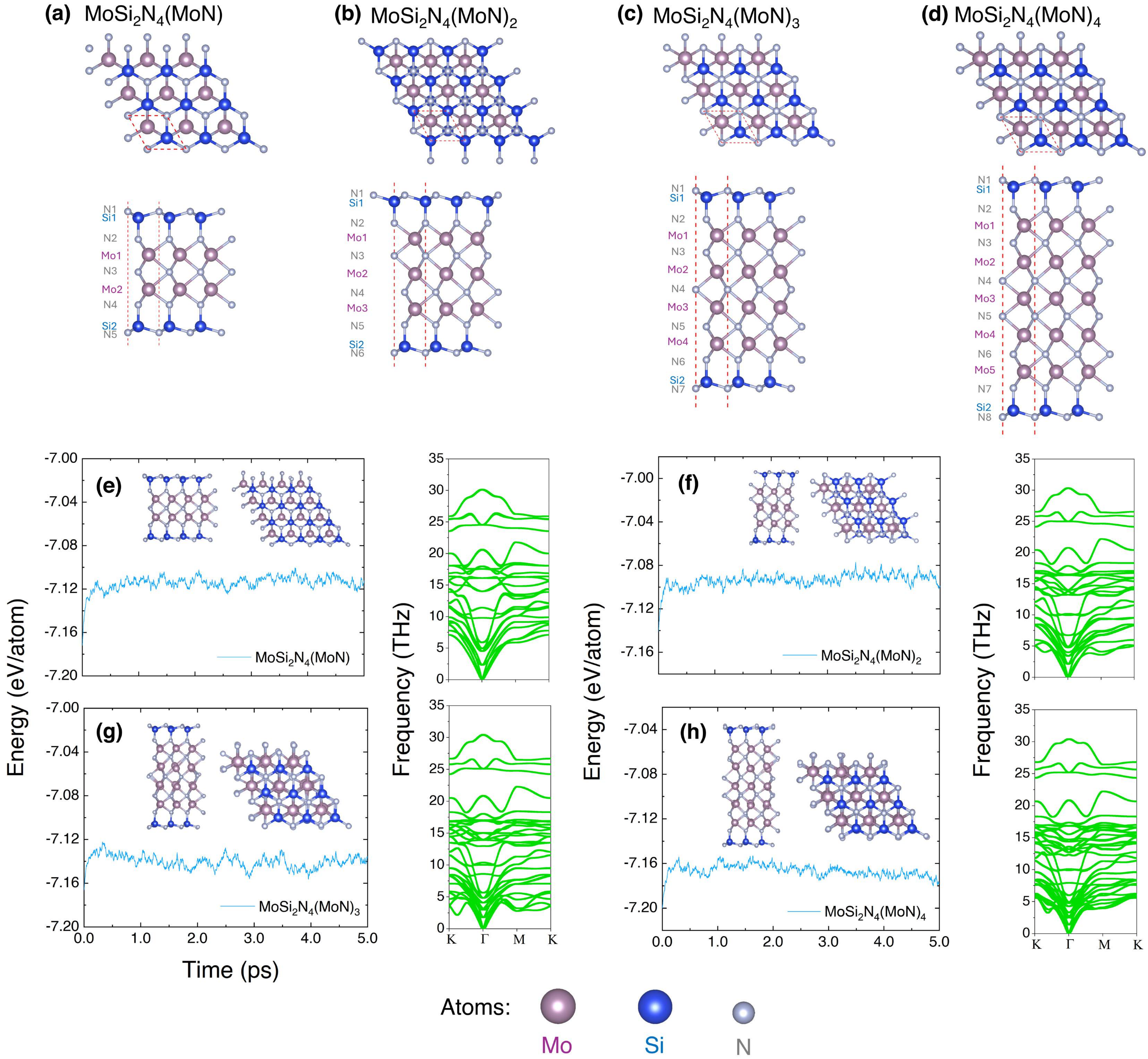}
\caption{\label{Fig2}\textbf{Lattice structure and stability tests.} (a-d) Top and side views of MoSi$_2$N$_4$(MoN)$_n$ monolayers. Red dotted lines denote the unit cells with the atoms being labelled. (e-h) Ab initio molecular dynamics simulation (AIMD) plots and phonon spectra showing the dynamic and thermodynamic stability of monolayers.}
\end{figure*}

We work with MoSi$_2$N$_4$(MoN)$_n$ in their most energetically favorable phases reported in the literature \cite{NSR}. The side and top views of the monolayers are shown in Fig. \ref{Fig2} (a-d). The monolayers possess out-of-plane hexagonal symmetry and consisted of overlapping 2H/1T-MoN$_2$ sublayers inserted between two outer $\alpha$-phase Si-N sublayers. The 2H- and 1T-MoN$_2$ phases are similar to the structural phases of 2H- and 1T-transition metal dichalcogenides (TMDCs), where one layer of N atoms are rotated by 180 degrees relative to the other layer of N atoms. The lattice constants and thickness ($t$) of the monolayers after geometrical optimization are given in Table \ref{Stability}, which are in close agreement with the values reported previously \cite{NSR}.

To evaluate the mechanical stability of the systems, elastic constants (C$_{i,j}$) values are obtained using VASPKIT \cite{Wang2021_vaspkit}, which implements the energy-strain method under the harmonic approximation:
\begin{equation}\label{Energy_Strain}
\Delta U = \frac{V_0}{2}\sum_{i=1}^{6}\sum_{j=1}^{6}C_{i,j}e_{i}e_{j}
\end{equation}
where $\Delta U$ is the energy difference between the strained and unstrained unit cell, $V_0$ is the volume of the unstrained unit cell and $e_i$($e_j$) is the matrix element of the strain vector. Due to the in-plane hexagonal symmetry of the studied systems, only three elastic constants ($C_{11}$, $C_{12}$, $C_{66}$ $= (C_{11}-C_{12})/2$) determines the material's mechanical properties. The 2D Young's modulus ($Y^{2D}$) and the Poisson's ratio ($\nu$) are calculated as $Y^{2D} = (C_{11}^2-C_{12}^2)/C_{11}$ and $\nu = C_{12}/C_{11}$, respectively. The 3D moduli of elasticity can allow us to make a comparison across materials of different thickness. As such, the values of $Y^{2D}$ are normalized by $t$ of the monolayers to obtain the 3D Young's Modulus ($Y^{3D}$), using $Y^{3D} = Y^{2D}/(t + 2t_{N, vdW})$, where $t_{N, vdW} = 1.55$ $\mathrm{\AA}$ is the vdW radius of N atom. The 3D bulk modulus ($K^{3D}$) and 3D shear modulus ($G^{3D}$) are calculated as $K^{3D} = Y^{3D}/(2-2\nu)$ and $G^{3D} = Y^{3D}/(2+2\nu)$, respectively [Table \ref{Stability}]. The calculated $Y^{3D}$ of MoSi$_2$N$_4$ (496 GPa) and MoSi$_2$N$_4$(MoN)$_n$ (450 $\sim$ 510 GPa) in this work are highly consistent with the experimentally obtained results \cite{Hong2020, NSR}. These $Y^{3D}$ values exceed those of other frequently studied 2D materials (e.g. TMDC: MoS$_2$ \cite{Cooper2013,Liu2014}, WS$_2$ \cite{Liu2014}; MXenes: Ti$_3$C$_2$T$_x$ \cite{Lipatov2018}, Nb$_4$C$_3$T$_x$ \cite{Lipatov2020}, Phosphorus-based 2D materials: Phosphorene \cite{khandelwal2017_phosphorene}, Phosphides \cite{Phosphides_GroupIII}), potentially making them more resistant to deformation when integrated with solid-state devices.

\begin{figure*}
\centering
\includegraphics[width=1\textwidth]{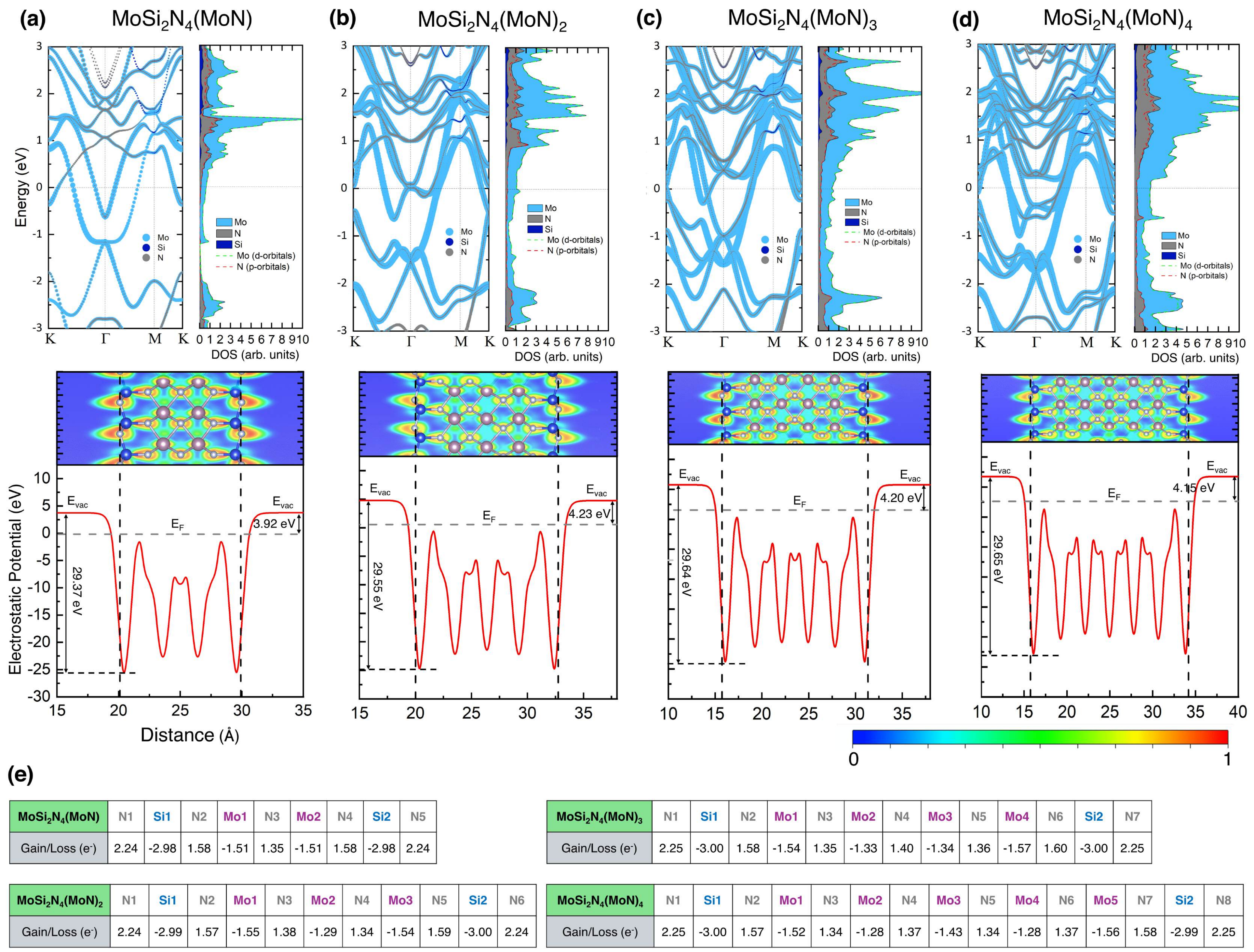}
\caption{\label{Fig3} \textbf{Electronic profiles of metallic monolayers.} (a-d) (Upper panel) Electronic band structure profiles of the metallic monolayers projected onto the atoms, with the atom-projected density of states (DOS) shown on the right. (Lower panel) Plane-averaged electrostatic potential and electron localization function (ELF) profiles. Color bar describes the normalized valence electron density with red(blue) denoting full electron localization(delocalization). (e) Tabulated values of electrons gain(loss) by each labelled atom obtained from Bader charge analysis.}
\end{figure*}

\subsubsection{Stability}
The elastic constants of all materials meet the Born-Huang’s requirement, thus confirming their mechanical stability (i.e. $C_{11}$ $>$ $C_{12}$ and $C_{66}$ $>$ 0). Thermodynamic stability are verified using Ab initio molecular dynamics simulation (AIMD). We choose the Andersen thermostat which couples each monolayer to a heat bath at a temperature of 300 K for 5 ps and set the Anderson probability of collision at 0.5. The atoms are shifted from their equilibriuim positions with no drastic changes to the interfacial characteristic with their neighbouring atoms during the period of collisions with random particles of the heat bath, thus confirming their thermodynamic stability. Dynamic stability of the monolayers are verified using PHONOPY program \cite{Phonopy_JPCM, Phonopy_JPSJ} under the finite difference method simulated on a supercell composed of 5 $\times$ 5 unit cells with a k-mesh sampling of 3 $\times$ 3 $\times$ 1. The absence of soft modes in the phonon spectra confirms the dynamical stability of the monolayers. The free energy versus time fluctuation of the lattice structures is calculated by AIMD, and their phonon spectra calculated by PHONOPY, are shown in Fig. \ref{Fig2}(e-h). 

The formation energy ($H^{sys}_{f}$) of a system can be calculated as:

\begin{equation}\label{formation_energy}
H^{sys}_{f} = E_{sys} - \sum_{i=1} n_{i}\mu_{i}
\end{equation}
where $E_{sys}$ is the total energy of the system, $n_i$ is the number of atoms of an element in the system and $\mu_{i}$ is the chemical potential of the element in its stable ground state phase. All the $H^{sys}_{f}$ are negative [Table \ref{Stability}], indicating the formation of the materials are energetically more favourable than the stable ground state phases of their constituent elements. These values are also comparable to those stable MA$_2$Z$_4$ monolayers reported in previous works, namely MoSi$_2$N$_4$ ($-0.955$ eV/atom); WSi$_2$N$_4$ ($-0.955$ eV/atom); MoGe$_2$N$_4$ ($-0.185$ eV/atom); and WGe$_2$N$_4$ ($-0.187$ eV/atom) \cite{Wang2021_MA2Z4}.

\begin{figure*}
\centering
\includegraphics[width=1\textwidth]{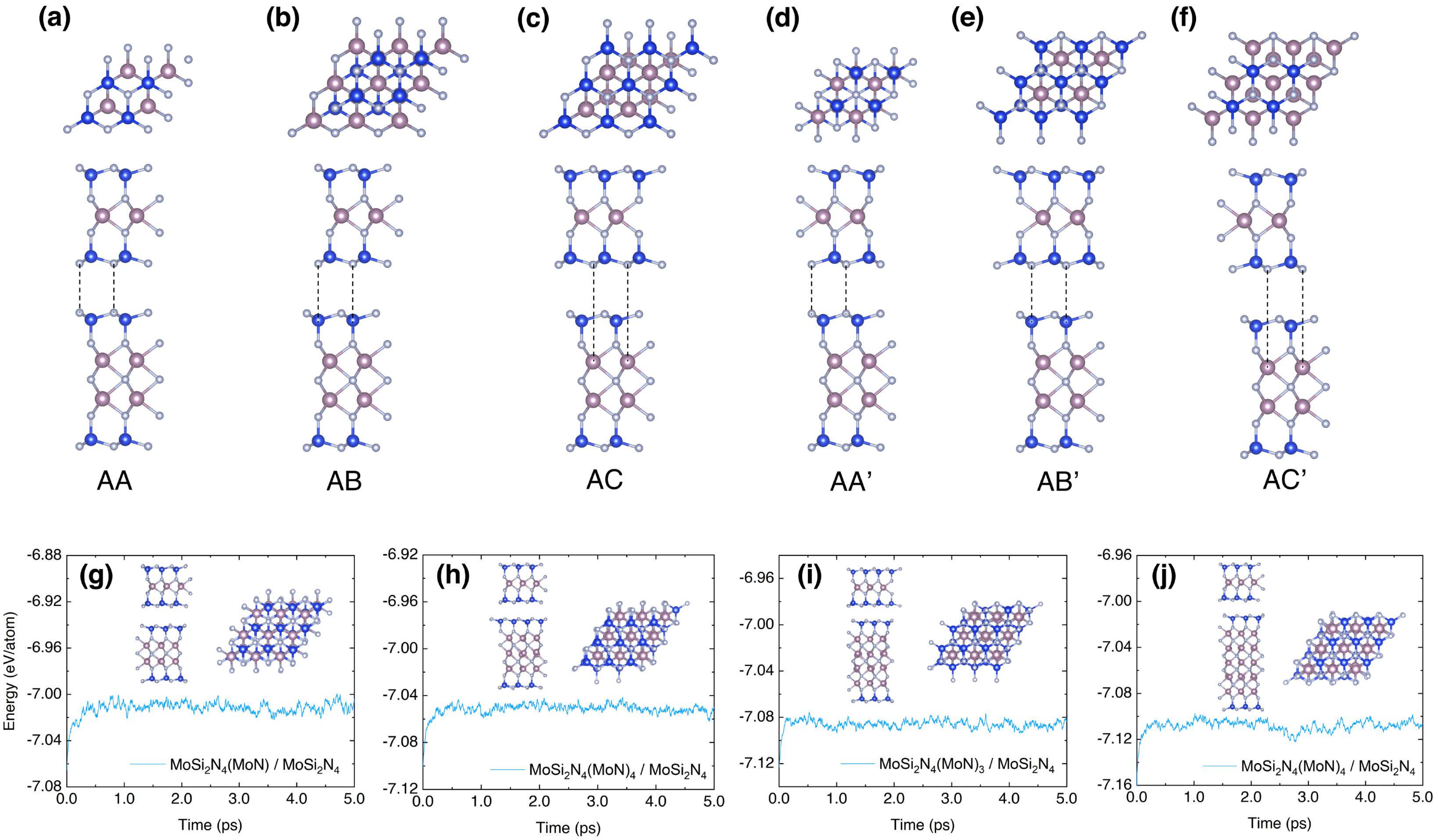}
\caption{\label{Fig4} \textbf{Stacking orders and AIMD plots of vdWHs.} (a-f) Top and side views of six different stacking orders of the van der Waals heterostructures (vdWHs) are being considered, using MoSi$_2$N$_4$/MoSi$_2$N$_4$(MoN) as an illustrative example. (g-j) AIMD plots of the AB' stacked vdWHs showing their thermodynamic stability after choosing the same thermostat parameters as the monolayers.}
\end{figure*}

\begin{figure*}
\centering
\includegraphics[width=1\textwidth]{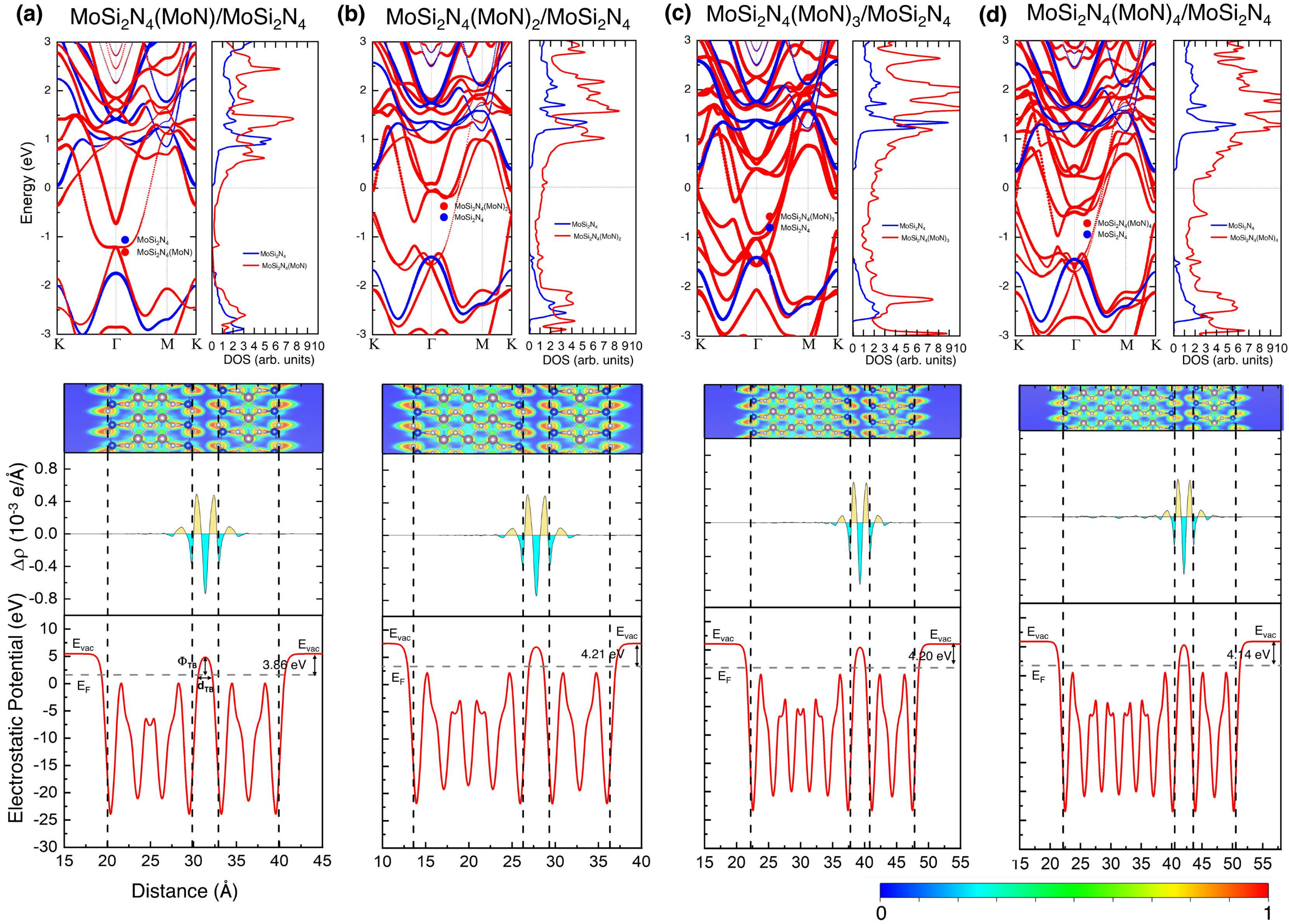}
\caption{\label{Fig5} \textbf{Electronic profiles of MoSi$_2$N$_4$/MoSi$_2$N$_4$(MoN)$_n$ vdWHs.} (a-d) (Upper panel) Electronic band structure of the metallic monolayers projected onto the constituent monolayers, with the layer-decomposed DOS shown on the right. (Lower panel) ELF profiles (top); plane-averaged charge density difference ($\Delta\rho$) profiles (middle); plane-averaged electrostatic potential profiles (bottom). Yellow(Cyan)-shaded regions in the $\Delta\rho$ profiles represents electron accumulation(depletion).}
\end{figure*}

\begin{figure*}
\centering
\includegraphics[width=1\textwidth]{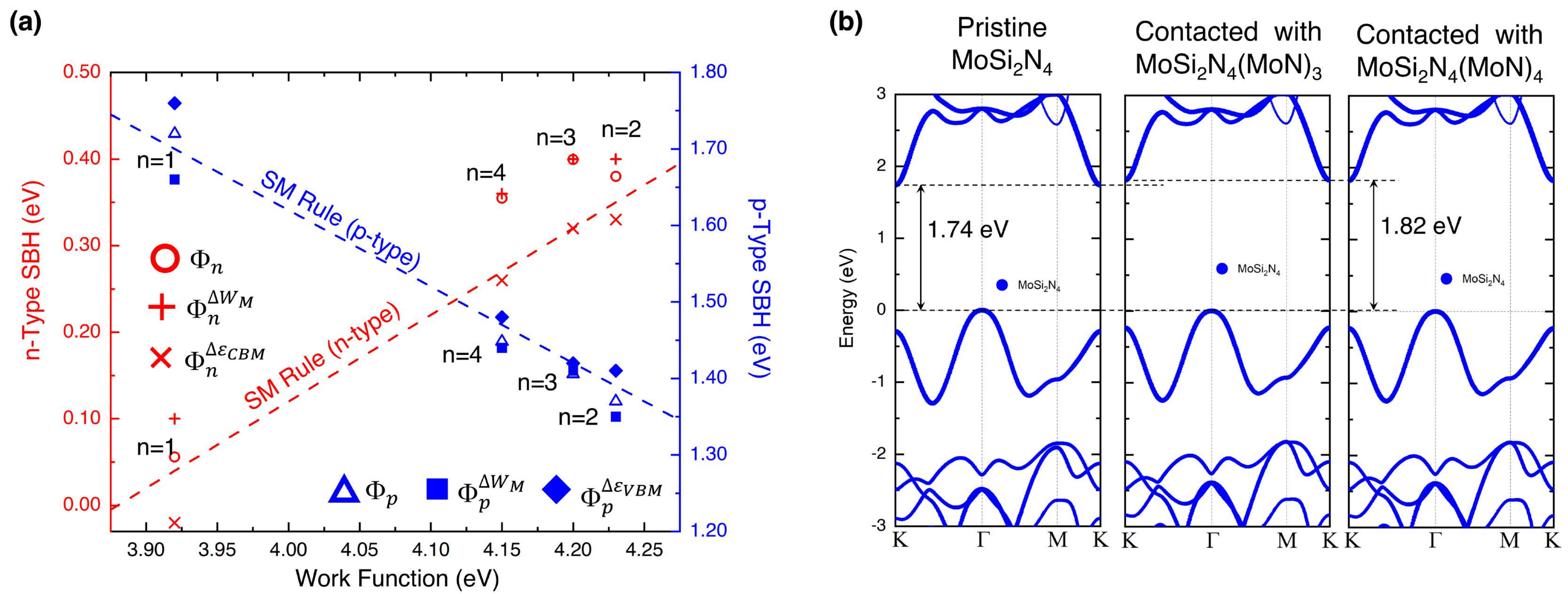}
\caption{\label{Fig6} \textbf{Deviation from the SM rule.} (a) n(p)-type SBH of MoSi$_2$N$_4$/MoSi$_2$N$_4$(MoN)$_n$ vdWHs. $\Phi_{n(p)}$, $\Phi^{\Delta W_M}_{n(p)}$, $\Phi^{\Delta \varepsilon_{CBM}}_{n}$ and $\Phi^{\Delta \varepsilon_{VBM}}_{p}$ denotes the n(p)-type SBH before any correction, the corrected SBH after accounting for the change in the metal's $W_M$, in the semiconductor's $\varepsilon_{CBM}$ and $\varepsilon_{VBM}$, respectively. (b) Electronic band structures of pristine MoSi$_2$N$_4$ (Left); MoSi$_2$N$_4$ that had been in contact with MoSi$_2$N$_4$(MoN)$_3$ (middle) and with MoSi$_2$N$_4$(MoN)$_4$ (right). }
\end{figure*}

\begin{table*}[t]
\centering
\caption{\label{Tunneling_SBH} Computed values of the tunneling barrier height ($\Phi_{TB}$) and width ($d_{TB}$), vdWH work function ($W_{vdWH}$), n(p)-type Schottky barrier height ($\Phi_{n(p)}$) of the vdWHs, as measured from the Fermi energy ($E_F$). $W_{vdWH}$ has the same value at two different sides of the vdWHs.} 
\begin{tabular}{>{\centering\arraybackslash}m{4cm}>{\centering\arraybackslash}m{1cm}>{\centering\arraybackslash}m{1cm}>{\centering\arraybackslash}m{1.5cm}>{\centering\arraybackslash}m{1.5cm}>{\centering\arraybackslash}m{1.5cm}} \hline \hline

\textbf{Heterostructures} & \textbf{$\Phi_{TB}$} & \textbf{$d_{TB}$}  & \textbf{$W_{vdWH}$} & $\Phi_{p}$ & $\Phi_{n}$ \\ \hline

MoSi$_2$N$_4$ / MoSi$_2$N$_4$(MoN) & 3.22 & 1.61 & 3.86 & 1.72 & 0.06 \\ 
MoSi$_2$N$_4$ / MoSi$_2$N$_4$(MoN)$_2$ & 3.56 & 1.68 & 4.21 & 1.37 & 0.38 \\  
MoSi$_2$N$_4$ / MoSi$_2$N$_4$(MoN)$_3$ & 3.57 & 1.69 & 4.20 & 1.41 & 0.40 \\ 
MoSi$_2$N$_4$ / MoSi$_2$N$_4$(MoN)$_4$ & 3.52 & 1.67 & 4.14 & 1.45 & 0.35 \\ \hline \hline

\end{tabular}%
\end{table*}





\subsubsection{\label{electronic_monolayer}Electronic properties}

The electronic band structures and the atom-projected density of states (DOS) of MoSi$_2$N$_4$(MoN)$_n$ are shown in Fig. \ref{Fig3}(a-d) (upper panel). All monolayers are metallic and the DOS are dominantly contributed by the d-orbitals of the transition metal atoms with some contributions from the p-orbitals of N atoms.

The electron localization function (ELF) of each monolayer can reveal the interfacial charge distribution among the atoms [bottom panel of Fig. \ref{Fig3}(a-d)]. ELF values of 1 (red) and 0 (blue) corresponds to perfect valence electron localization and delocalization, respectively. The valence electron distribution is highly localised around N but less around Mo, suggesting the ionic nature of the bonding within the inner Mo-N sublayer. Additionally, there is almost no valence electrons localised around Si, resulting in the outer ionic Si-N bonds being stronger than the inner Mo-N bonds. This behavior is also evident in the Bader charge analysis [Fig. \ref{Fig3}(e)], which shows a higher amount of electron gain by the outer N (2.2 electrons) as compared to the inner N (1.3 - 1.6 electrons) for all MoSi$_2$N$_4$(MoN)$_n$ monolayers. We further calculate the plane-averaged electrostatic potential along the out-of-plane direction of MoSi$_2$N$_4$(MoN)$_n$ [lower panel of Fig. \ref{Fig3}(a-d)], which allows us to extract the $W_M$ for these metallic monolayers (3.92, 4.23,  4.20 and 4.15 eV for $n$ = 1, 2, 3 and 4, respectively).



\begin{table*}[t]
\centering
\caption{\label{Stacking_Energy} Binding energy ($E_{B}$) of the vdWHs for each stacking order. The unit is in meV per atom.} 
\begin{tabular}{>{\centering\arraybackslash}m{5cm}>{\centering\arraybackslash}m{2cm}>{\centering\arraybackslash}m{2cm}>{\centering\arraybackslash}m{1.5cm}>{\centering\arraybackslash}m{1.5cm}>{\centering\arraybackslash}m{1.5cm}>{\centering\arraybackslash}m{2cm}}
\hline \hline
\textbf{Materials} & \textbf{AA} & \textbf{AB} & \textbf{AC} & \textbf{AA'} & \textbf{AB'} & \textbf{AC'}  \\ \hline 
MoSi$_2$N$_4$ / MoSi$_2$N$_4$(MoN)   &   -8.44   &  -11.27 &  -11.26 & -8.58 &   -12.08  & -10.34  \\ 
MoSi$_2$N$_4$ / MoSi$_2$N$_4$(MoN)$_2$   &  -7.54 &    -10.03 &  -10.02 &  -7.64   & -10.76  & -9.21 \\ 
MoSi$_2$N$_4$ / MoSi$_2$N$_4$(MoN)$_3$  &  -6.83 & -9.05  & -9.04 & -6.83  & -9.74 &  -8.31 \\ 
MoSi$_2$N$_4$ / MoSi$_2$N$_4$(MoN)$_4$  &  -6.16 & -8.23  & -8.23  & -5.99  & -8.83  & -7.56 \\ \hline \hline
\end{tabular}%
\end{table*}

\subsection{\label{vdWH} MoSi$_2$N$_4$/MoSi$_2$N$_4$(MoN)$_n$ vdWHs} 

Based on the alignment of the metal $W_M$ with the semiconductor band edges, the contact type of MS vdWH will either be Ohmic or Schottky. In the following discussion of the vdWHs constituting of monolayer MoSi$_2$N$_4$ and MoSi$_2$N$_4$(MoN)$_n$, our results are organised into four subsections: \emph{1. Stability}; \emph{2. Electronic properties}; \emph{3. Zero-Dipole contact}; and \emph{4. Pressure sensing}.  

\subsubsection{Stability}
We consider six different stacking orders [Fig. \ref{Fig4}(a-f) shows MoSi$_2$N$_4$(MoN)/MoSi$_2$N$_4$ as an illustrative example] and calculated their binding energy ($E_{B}$) using:

\begin{equation}\label{binding_energy}
E_{B} = (E_{vdWH} - E_{MoSi_{2}N_{4}} - E_{MoSi_{2}N_{4}(MoN)_{n}})/N
\end{equation}

\begin{figure*}
\centering
\includegraphics[width=1\textwidth]{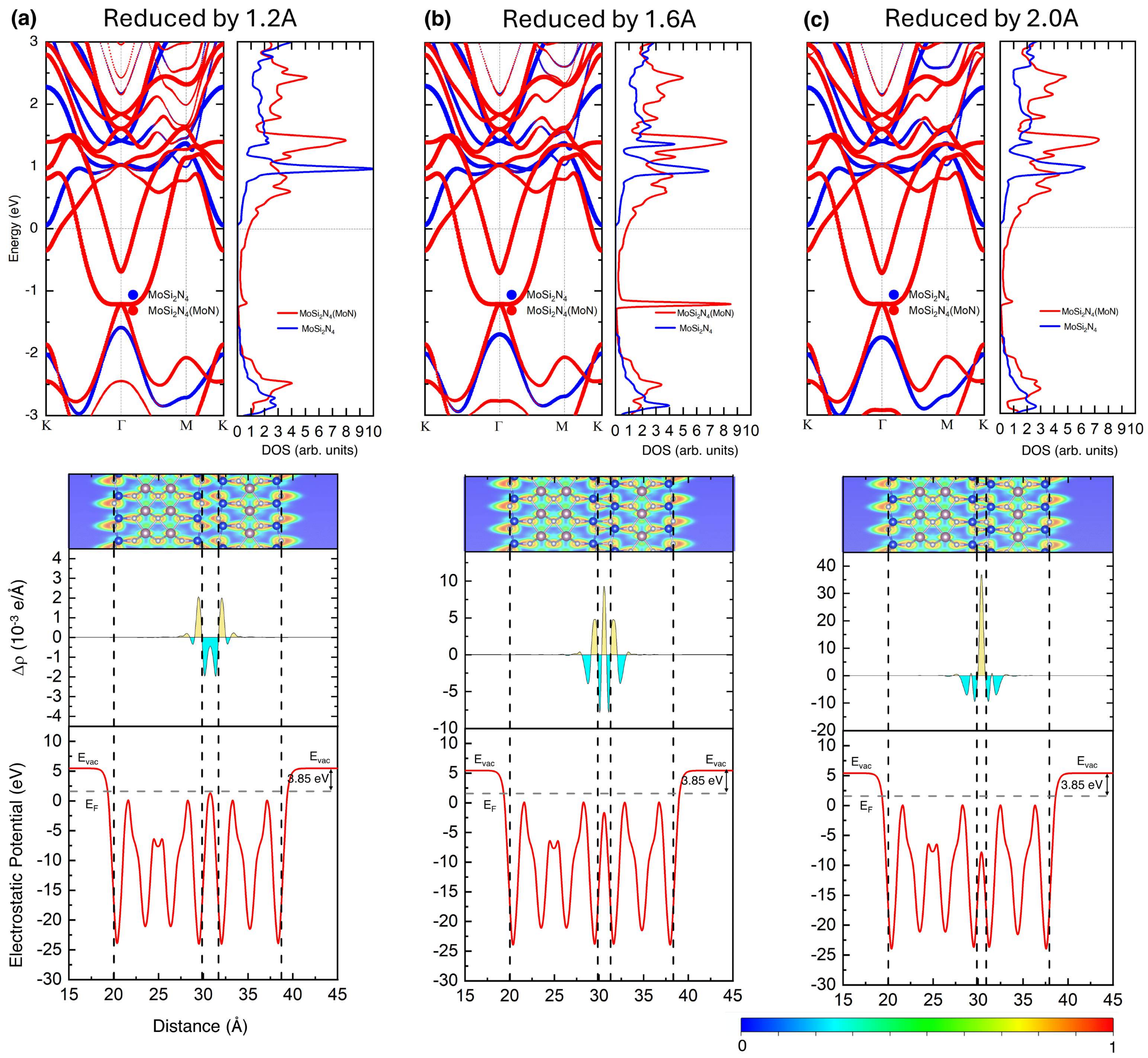}
\caption{\label{Fig7} \textbf{Interfacial charge redistribution.} (a-c) (Upper panel) Electronic band structures; (Lower panel) ELF profiles (top); $\Delta\rho$ profiles (middle); plane-averaged electrostatic potential profiles (bottom) of MoSi$_2$N$_4$/MoSi$_2$N$_4$(MoN) vdWH when the interlayer distance is reduced by (a) 1.2 $\mathrm{\AA}$; (b) 1.6 $\mathrm{\AA}$; and (c) 2.0 $\mathrm{\AA}$.}
\end{figure*}

where $N$, $E_{vdWH}$, $E_{MoSi_2N_4}$ and $E_{MoSi_2N_4(MoN)_n}$ is the number of atoms, the energy of the vdWH, contacted MoSi$_2$N$_4$ and contacted MoSi$_2$N$_4$(MoN)$_n$, respectively. The $E_B$ values are negative, which indicates that the formation of the vdWHs from their constituent monolayers are energetically favourable [Table \ref{Stacking_Energy}]. AB' stacking order is found to have the largest $|E_{B}|$ for all the vdWHs and thus we only worked with this stacking order.

The vertical interlayer distances ($d$) measured between the interface N atoms, are found to be about 3 $\mathrm{\AA}$ [Table \ref{Stability}], which are slightly less than the interlayer distance in most other vdWHs reported in literature \cite{Pei2023}. The vdWHs retain the hexagonal symmetry of their constituent monolayers and are mechanically stable, with their $Y^{3D}$ being comparable to the monolayers despite having larger $Y^{2D}$ [Table \ref{Stability}]. AIMD simulations using the same parameters as the monolayers [Fig. \ref{Fig4}(g-j)] show that the vdWHs are thermodynamically stable, whereas the phonon calculations of the vdWHs are not conducted due to computational resource constraints. It is reasonable to assume that the vdWHs are also dynamically stable, given the stability of the monolayers and the negative $E_{B}$ upon forming the vdWHs.

\subsubsection{Electronic properties}
The contact type of MoSi$_2$N$_4$/MoSi$_2$N$_4$(MoN)$_n$ vdWHs can be revealed from their layer-decomposed electronic band structure and layer-decomposed DOS [upper panel of Fig. \ref{Fig5}(a-d)] profiles. Intriguingly, the n(p)-type SBH values [Table \ref{Tunneling_SBH}] lie close to those predicted by the SM rule, which are 0.04(1.70), 0.35(1.39), 0.32(1.42) and 0.27(1.47) eV for $n$ = 1, 2, 3 and 4, respectively. Figure \ref{Fig5}(a-d) (lower panel) shows the ELF, plane-averaged charge density difference ($\Delta\rho$) and electrostatic potential profiles of the vdWHs. $\Delta\rho$ is calculated as $\Delta\rho = \rho_{vdWH}-\rho_{MoSi_2N_4}-\rho_{MoSi_2N_4(MoN)_n}$, whereby $\rho_{vdWH}$, $\rho_{MoSi_2N_4}$ and $\rho_{MoSi_2N_4(MoN)_n}$ is the charge density of the vdWH, MoSi$_2$N$_4$ and MoSi$_2$N$_4$(MoN)$_n$, respectively. Despite the vdWHs not having an out-of-plane mirror symmetry, the vdWH work function ($W_{vdWH}$) at both ends of the vdWH are similar [Table \ref{Tunneling_SBH}], which strongly points to an absence of charge transfer across the interface. This is further shown in the nearly symmetrical $\Delta\rho$ profiles, where electrons accumulate at the interfacial N atoms but deplete at the midpoint of the vdW gap. ELF profiles show the shorter span of the electron density localised at the interfacial N atoms, compared to the longer span at both ends of the vdWHs. Bader charge analysis further confirms a negligible amount of electrons ($<$ 0.001 electrons) are being transferred across the vdWHs. The absence of charge transfer at the interface is peculiar for vdWHs that are constructed using different monolayers, since the broken out-of-plane symmetry should induce a built-in electric field across the vdWH. In our cases, the near absence of charge transfer is attributed to similar electronegativity of the constituent Si and N atoms that make up the Si-N sublayer at the surfaces. Moreover, the wave function of the electrons at the vdWH interface overlaps strongly at close distances, whereby electrons with parallel spins in the anti-symmetric wave function will be kept further apart via Pauli repulsion. This causes the electrons at the interface to be `push-back' towards the constituent monolayers instead of being transferred across.

Deviation from the SM rule in MoSi$_2$N$_4$/MoSi$_2$N$_4$(MoN)$_n$ vdWHs, is attributed to the combined changes in the CBM ($\Delta \varepsilon_{CBM}$) and VBM ($\Delta \varepsilon_{VBM}$) of MoSi$_2$N$_4$, and change in the work function ($\Delta W_M$) of MoSi$_2$N$_4$(MoN)$_n$ upon forming the vdWHs. Figure \ref{Fig6}(a) shows the n(p)-type SBH obtained from the vdWHs through DFT calculations, and the n(p)-type SBH values ($\Phi^{\Delta W_M}_{n(p)}$, $\Phi^{\Delta \varepsilon_{CBM}}_{n}$ and $\Phi^{\Delta \varepsilon_{VBM}}_{p}$) obtained after taking into account these changes. By accounting for these corrections, the SBH determined from SM rule ($\Phi_{n(p),SM}$) can be recovered using:

\begin{equation}\label{SBH_n}
\Phi_{n,SM} = \Phi_{n}+\Delta\varepsilon_{vac}+\Delta \varepsilon_{CBM} - \Delta W_{M}
\end{equation}

for n-type SBH, and using

\begin{equation}\label{SBH_p}
\Phi_{p,SM} = \Phi_{p}-\Delta\varepsilon_{vac}-\Delta \varepsilon_{VBM} + \Delta W_{M}
\end{equation}

for p-type SBH. $\Delta\varepsilon_{vac}$ is the vacuum level offset induced by the interface dipole which is zero in all of our vdWHs. We note that the deviation of n-type SBH from the SM rule is more pronounced for  $n$ = 3, 4 as seen in Fig. \ref{Fig6}(a). Figure \ref{Fig6}(b) shows the electronic band structure of the contacted MoSi$_2$N$_4$ for $n$ = 3, 4 which reveal the band gap of the contacted MoSi$_2$N$_4$ being slightly larger than the case of pristine MoSi$_2$N$_4$. Therefore for $n$ = 3, 4, the considerable deviation of the n-type SBH from the SM rule, is mainly attributed to the rearrangement of MoSi$_2$N$_4$ surface atoms when they are contacted with metal, suggesting that the band gap of MoSi$_2$N$_4$ is sensitive to surface couplings even in the absence of lattice strain.


\subsubsection{Zero-Dipole contact}
We further investigate the interfacial characteristic of MoSi$_2$N$_4$/MoSi$_2$N$_4$(MoN) vdWH due to the atypical $\Delta\rho$ profiles. MoSi$_2$N$_4$/MoSi$_2$N$_4$(MoN) vdWH has the lowest n-type SBH among the studied vdWHs and thus is of particular interest due to its potential for forming MS contacts with low resistance. To increase the effects of surface couplings, we decrease $d$ in MoSi$_2$N$_4$/MoSi$_2$N$_4$(MoN) vdWH by 1.2, 1.6 and 2.0 $\mathrm{\AA}$. The out-of-plane positions of the surface atoms are held fixed while we allow the other atoms of the vdWH to relax in all directions. Intriguingly, at all three close-contact distances, the electronic band structures and SBH values remain unchanged. Interface dipole remains absent which implies that there is practically no net charge transfer between the constituent monolayers. 

An analysis on the interfacial electronic characteristic shows that when $d$ is decreased by 1.2 $\mathrm{\AA}$, the tunneling barrier at the vdW gap vanishes, which is desirable in reducing the contact resistance in MS contacts. The `push-back' effect becomes more prominent, as seen from the higher peaks of the electron accumulation regions in Fig. \ref{Fig7}(a) compared to Fig. \ref{Fig5}(a), whereby the electron accumulation regions are highly localised at the Si-N sublayer while the depletion region extends across the vdW gap. When $d$ is reduced by 1.6 $\mathrm{\AA}$, the charge distribution remains symmetrical but electrons start to accumulate at the midpoint of the vdW gap while the depletion regions form at the inner Si-N bonds, as seen in Fig. \ref{Fig7}(b). When the interlayer distance is reduced by 2.0 $\mathrm{\AA}$, the electron accumulation region extends across the vdW gap and the `push-back' effect vanishes. This kind of interfacial charge redistribution, whereby the electrons are symmetrically shared between two constituent materials, is also reported in Mo$_2$CF$_2$/MoS$_2$ \cite{Quasicovalent} vdWH and layered PtS$_2$ \cite{PtS2_covalent}. Here, although the interface distance is less than the sum of the covalent radius of the surface atoms ($\sim$ 1.82 $\mathrm{\AA}$), we remark that the interfacial characteristic in MoSi$_2$N$_4$/MoSi$_2$N$_4$(MoN) vdWH at such close-contact regime is not chemical in nature but is of \emph{quasi-bonding}. Gap states in MS contacts, which are signature of chemical bonds formed through the hybridization of orbitals between the metal and semiconductor \cite{jinglu2021_schottky}, is absent in the electronic band structures [top panel of Fig. \ref{Fig7}(a-c)].             

\begin{figure}[t]
\includegraphics[width=0.485\textwidth]{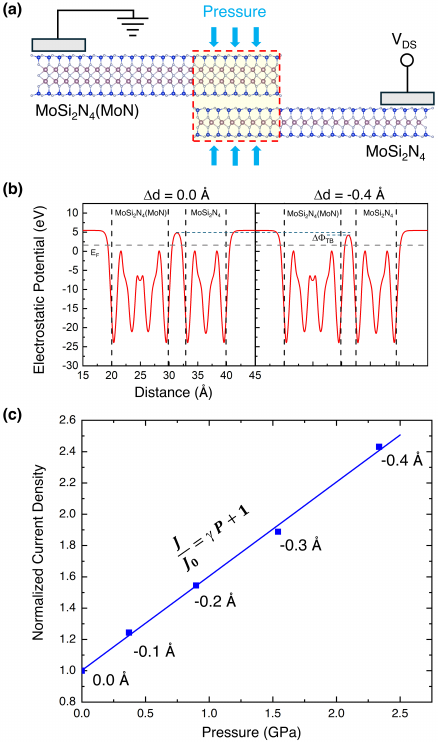}
\caption{\label{Fig8}\textbf{MoSi$_2$N$_4$/MoSi$_2$N$_4$(MoN) vdWH for pressure sensing.} (a) Schematic of the set-up. Red dotted box indicates the region where electrons tunnel from MoSi$_2$N$_4$(MoN) to MoSi$_2$N$_4$. The current direction is indicated by black arrows. (b) Electrostatic potential profile showing that the reduction of interlayer distance will cause a reduction to the tunneling barrier. (c) Normalized current density as a function of the external pressure applied vertically on the vdWH. We set $V_{SD}$ = 0.2 V.}
\end{figure} 

\subsubsection{Pressure sensing}


The modulation of electrical current via mechanical stimuli forms the basis of piezoelectronic transistors and pressure sensors \cite{qin2024_pressure_sensor, wu2016_piezotronics}. For vdWHs, the interlayer van der Waals (vdW) gap between the constituent materials produce an electrostatic potential barrier [see Table \ref{Tunneling_SBH} and bottom panel of Fig. \ref{Fig5}(a-d)] that sensitively influence the electron tunneling and transport through an vdWH. As mechanical pressure can modulate the vdW gap \cite{pei2022_high_pressure} which can subsequently lead to a modulation of the electron transport current across the contact interface, we propose that such mechanism can be used achieve pressure sensing using the MoSi$_2$N$_4$/MoSi$_2$N$_4$(MoN) vdWH [see Fig. 8(a) for schematic of the device structure]. 
MoSi$_2$N$_4$/MoSi$_2$N$_4$(MoN) vdWH has two aspects that are potentially beneficial pressure sensing: (1) The ultralow Schottky barrier of MoSi$_2$N$_4$/MoSi$_2$N$_4$(MoN) can facilitate sizable electrical current flow; and, more importantly, (2) the zero-dipole nature of MoSi$_2$N$_4$/MoSi$_2$N$_4$(MoN) suggests that the vdW gap tunneling barrier modulation will be predominantly caused by the interlayer distance variation [Fig. \ref{Fig8}(b)] from the equilibrium value (defined as $\Delta d$) without being additionally complicated by an interface dipole.

We consider the interlayer distance $d$ is changed by $\Delta d$ when a force $F$ is uniformly exerted on an area $A$. In the linear response regime, work done by $F$ in reducing the interlayer distance by $\Delta d$ leads to an energy change $\Delta E_{sys}$ of the system. As the applied pressure is related to $F$ by $P = F/A$ where $A$ is the area, the following expression can be obtained \cite{nguyen2022_C3N4}:
\begin{equation}\label{pressure}
P = \frac{\Delta E_{sys}}{A|\Delta d|}
\end{equation}
Here $\Delta E_{sys}$ per unit cell can be obtained directly from the DFT calculations of the equilibrium and strained vdWHs, and the $A \approx 7.33 \mathrm{\AA}^2$ is the area of the unit cell. 

We can now construct a transport model for a pressure sensing device [Fig. \ref{Fig8} (a)]. 
The tunneling probability ($\mathcal{T}(\varepsilon_{\bf{k}})$) across the vdW gap for an incident electron with energy $\varepsilon_{\bf{k}}$ is \cite{Simmons1963}:  
%
%
%
\begin{equation}\label{tunneling_probability}
\mathcal{T}(\varepsilon_{\bf{k}})=\exp\left(\frac{-2}{\hbar}\int_{\text{barrier 
 width}}\sqrt{2m_{e}(V(x)-\varepsilon_{\bf{k}})} \, dx\right)
\end{equation}
where  $\hbar$ is the reduced Planck constant, $m_e$ is the mass of the free electron and $x$ is the distance along the vdW gap. The $V(x)$ is the vdW tunneling barrier. Instead of assuming a simplified square barrier profile, we can extract $V(x)$ \emph{fully numerically} from the DFT calculated electrostatic potential profile [Figs. 8(b)] for various interlayer distance. 
In addition, the tunneling barrier is further modulated by an applied bias voltage $V_{DS}$, where such effect is explicitly included in $V(x)$ \cite{Simmons1963}. The electrical current flowing from the metal into the semiconductor ($J_{m\rightarrow s}$) is:
%
%
%
%
%
\begin{multline} \label{m_to_sc}
J_{m \rightarrow s} = \frac{ge}{(2\pi)^{2}} \int \mathbf{v}_{tr} f_{m} \left( 1 - f_{s}\right) \mathcal{T} \, d^{2}\mathbf{k} \\
= \frac{e}{\hbar} \int_{\Phi_{n}}^{\infty} \frac{\partial \varepsilon_{\mathbf{k}}}{\partial k_{tr}} f_{m}\left( 1 - f_{s}\right) \mathcal{T} D_{s} \, d\varepsilon_{\mathbf{k}}
\end{multline}
where $g$ is the spin degeneracy of an electron, $e$ is the charge of an electron, $\mathbf{k}$ is the wave vector of an electron, $k_{tr}$ is the wave vector of an electron in the transport direction taken along $K \rightarrow \Gamma$ path, $\mathbf{v}_{tr}$ is the velocity of an electron in the transport direction, $f_{m}$ and $f_{s}$ is the carrier distribution of electrons in metal and semiconductor, respectively. We consider $f_{m} = f_0(\varepsilon_k, E_F, T)$ and $f_{s} = f_0(\varepsilon_k +eV_{DS}, E_F, T)$ where $f_0$ is the equilibrium Fermi-Dirac distribution function and $T = 300$ K at room temperature. 

As the Fermi surface of the semiconductor is smaller than that of the metal [see band structure in Fig. 5(a)], the carrier conduction is limited by the DOS in the semiconductor ($D_{s}$): 
\begin{equation}
    D_{s}(\varepsilon_{\bf{k}}) \, d\varepsilon_{\mathbf{k}} = \frac{g}{2\pi} k \, dk
\end{equation}
For the electrical current flowing from the semiconductor into the metal ($J_{s \rightarrow m}$):   
%
%
%
%
%
\begin{equation} \label{sc_to_m}
J_{s \rightarrow m} = \frac{e}{\hbar} \int_{\Phi_{n}}^{\infty} \frac{\partial \varepsilon_{\bf{k}}}{\partial k_{tr}} f_{s} \left(1 - f_{m}\right) \mathcal{T}D_{s} \, d\varepsilon_{\bf{k}}
\end{equation}
Here the integration starts from $\Phi_n$ to account for the presence of a SBH. The net electrical current density ($J$) is then obtained as $J = J_{m\to s} - J_{s\to m}$, which yields:

%
%
\begin{equation} \label{net_m_to_sc}
J = \frac{e}{\hbar} \int_{\Phi_{n}}^{\infty} \frac{\partial \varepsilon_{\bf{k}}}{\partial k_{tr}} \left( f_{m} - f_{s}\right) \mathcal{T} D_{s} \, d\varepsilon_{\bf{k}}
\end{equation}
%
%
%
Generally $J$ is in the order of $1$ mA $\mu$m$^{-1}$ under $V_{DS} = 0.2$ V, which are well within the detection capability of probing apparatus \cite{mitta2020_electrical}. The current densities normalized by the unstrained vdWH ($J_{0}$ = 1.663 mA $\mu$m$^{-1}$) are plotted in Fig. \ref{Fig8}(c) against pressure. The current-pressure curve exhibit exceptional linearity and can be well fitted by with a slope parameter of $\gamma = 0.6$ GPa$^{-1}$. Such linearity suggests the potenital and practicality of the MoSi$_2$N$_4$(MoN)/MoSi$_2$N$_4$ contact heterostructure as a pressure sensor.

\section{\label{sec:Conclusion}Conclusion}

In summary, we investigated the electronic properties of the most stable phases of MoSi$_2$N$_4$(MoN)$_n$ ($n = 1-4$) and their vdWHs formed with MoSi$_2$N$_4$ based on DFT calculations. The work function of the metals are extracted from the plane-averaged electrostatic potential profiles and Bader charge analysis was performed to quantitatively uncover the electron distribution at each atom. 


As for the vdWHs, we showed that n-type Schottky contact is formed in all of our cases, with MoSi$_2$N$_4$/MoSi$_2$N$_4$(MoN) vdWH exhibiting the lowest n-type SBH. Intriguingly, a \emph{zero-dipole} contact is formed across the vdWH and the `push-back' effect causes the charge redistribution to be nearly symmetrical. Such \emph{zero-dipole} MS contact suggests that the scope of our findings may be extended to vdWHs that are formed by contacting homologous compounds with their blueprint materials, in which no net charge transfer will occur between the constituent monolayers. As a result of the absence of the interface dipole, the Schottky slope parameter of MoSi$_2$N$_4$ when in contact with MoSi$_2$N$_4$(MoN)$_n$, lies close to the ideal SM slope value of 1, a result that has been shown in our earlier work \cite{Tho2022}.

Furthermore, we identified MoSi$_2$N$_4$/MoSi$_2$N$_4$(MoN) vdWH as a potentially low-resistance and robust MS contact by varying the interlayer distance. The electronic band structure of MoSi$_2$N$_4$/MoSi$_2$N$_4$(MoN) vdWH remains unaltered even at close-contact distances where surface interactions are strong, with the SBH being insensitive to interlayer distance changes. These results demonstrate the robustness of the SBH in MoSi$_2$N$_4$/MoSi$_2$N$_4$(MoN) vdWH against external perturbations that changes interlayer distance, further suggesting that similar properties might be present for other vdWHs that are constructed from homologous compounds and their blueprint materials. Finally, based on the promising interfacial characteristics of  MoSi$_2$N$_4$/MoSi$_2$N$_4$(MoN) vdWH, we investigated the vdWH's potential to be used as a pressure sensor, and found its tunneling current density output to display good linearity with pressure. We anticipate that our findings on the robustness of \emph{zero-dipole} Schottky contact will reveal new pathways toward advancing the MS contacts for nanodevice designs.

\section*{CRediT authorship contribution statement}
\textbf{Che Chen Tho}: Data curation (lead); Formal analysis (lead);
Investigation (equal); Writing – original draft (equal); Visualization (lead); Writing – review \& editing (supporting);
\textbf{Yee Sin Ang}: Conceptualization (lead); Investigation (equal); Formal analysis (supporting); Supervision (lead); Funding acquisition (lead); Visualization (supporting); Writing – original draft (equal); Writing – review \& editing (lead).

\section*{Declaration of competing interest}
The authors declare that they have no known competing financial interests or personal relationships that could have appeared to influence the work reported in this paper.

\section*{Data availability}

Data will be made available on request.

\section*{\label{sec:Acknowledgement}Acknowledgement}
This work is funded by the Singapore Ministry of Education (MOE) Academic Research Fund (AcRF) Tier 2 Grant (MOE-T2EP50221-0019) and SUTD-ZJU IDEA Thematic Research Grant (SUTD-ZJU (TR) 202203). The computational work for this article was performed on resources of the National Supercomputing Centre, Singapore (https://www.nscc.sg).

\section*{Appendix A. Supplementary Data}
Supporting Information is available from the Wiley Online Library or from the author.


\providecommand{\noopsort}[1]{}\providecommand{\singleletter}[1]{#1}%

\end{document}